\documentclass[11pt]{iopart}
\usepackage{graphicx}
\usepackage{amssymb}
\usepackage{iopams}
\usepackage{amsthm}
\usepackage{latexsym}
\usepackage{hyperref}
\usepackage{fontenc}
\usepackage{geometry}
\usepackage{setspace}

\newcommand{\bdm}{\begin{displaymath}}
\newcommand{\edm}{\end{displaymath}}
\newcommand{\be}{\begin{equation}}
\newcommand{\ee}{\end{equation}}
\newcommand{\bea}{\begin{eqnarray}}

\newcommand{\eea}{\end{eqnarray}}

\begin{document}

\title[Fokker-Planck-Rosenbluth-Type Equations in 1PN Approximation]
{Fokker-Planck-Rosenbluth-Type Equations for Self-gravitating Systems in 1PN
Approximation}

\author{Javier Ramos-Caro\footnote[1]{e-mail:javiramos1976@gmail.com}
and Guillermo A. Gonz\'alez\footnote[2]{e-mail:gonzalez@gag-girg-uis.net}}

\address{Escuela de F\'isica, Universidad Industrial de Santander, A. A. 678,
Bucaramanga, Colombia}

\begin{abstract}
We present two formulations of Fokker-Planck-Rosenbluth-type (FPR)
equations for many-particle self-gravitating systems, with first
order relativistic corrections in the post-Newtonian approach (1PN).
The first starts from a covariant Fokker-Planck equation for a
simple gas, introduced recently by G. Chac\'on-Acosta and G. Kremer
\cite{chacon}. The second derivation is based on the establishment
of an 1PN-BBGKY hierarchy, developed systematically from the 1PN
microscopic law of force and using the Klimontovich-Dupree (KD)
method. We close the hierarchy by the introduction of a two-point
correlation function that describes adequately the relaxation
process. This picture reveals an aspect that is not considered in
the first formulation: the contribution of ternary correlation
patterns to the diffusion coefficients, as a consequence of the
nature of 1PN interaction. Both formulations can be considered as a
generalization of the equation derived by Rezania and Sobouti in
2000 \cite{rezania}, to stellar systems where the relativistic
effects of gravitation play a significant role.
\end{abstract}


\section{Introduction \label{intro}}

The collisionless Boltzmann equation in 1PN approximation for a self
gravitating system, imbedded in an otherwise flat space-time, was
derived by Rezania and Sobouti in 2000 \cite{rezania}, finding some
relevant solutions. The aim of the present paper is to go beyond the
collisionless case and incorporate situations where encounters
between particles play a significant role, and derive tractable
kinetic equations describing the evolution of this class of self
gravitating systems. As it was pointed out by Kandrup, the
astrophysical objects where the relativistic effects could play a
significant role in the relaxation process are the galactic nucleus
and, perhaps, the relativistic star clusters with age shorter than
its relaxation time \cite{kand3}. There is solid observational
evidence that there exists nuclei containing massive black holes
formed due to the dynamical instability of relativistic systems of
stars. In particular, the relativistic effects might be important in
the case of a galactic nucleus decoupled from the remainder of the
galaxy, which becomes even more dense (and relativistic) at a time
scale of the order of its relaxation time \cite{rees,shapiro}. Since
these systems are composed by stars with velocity much less than the
speed of light, the post-Newtonian approach supplies the adequate
tool to investigate their behavior.

It is well known that the evolution of typical globular clusters,
where it is assumed that relativistic corrections are unimportant,
is described satisfactorily by the Fokker-Planck kinetic equation.
In the local approximation, it is possible obtain explicit relations
for the diffusion coefficients, in terms of Rosenbluth potentials,
and such equation takes a very tractable form \cite{BinTre} (in this
paper we call it as the FPR equation):
\begin{equation}\label{fokk-planck}
\frac{\partial f}{\partial t} + {\bf v} \cdot \frac{\partial
f}{\partial {\bf x}} + {\bf g} \cdot \frac{\partial f}{\partial {\bf
v}} = -\frac{\partial}{\partial
   v_{i}}(A^{i}_{N}f)+\frac{1}{2}\frac{\partial^{2}}{\partial v_{i}\partial
   v_{j}}(B^{ij}_{N}f),
\end{equation}
where we have used the sum convention ($i,j=1,2,3$). Here
$A^{i}_{N}$ and $B^{ij}_{N}$ are the diffusion coefficients or
Rosenbluth's coefficients (the subscript $N$ remarks the fact that
we are dealing with the classical Newtonian theory). Such equation
describes the evolution of the distribution function $f({\bf x},{\bf
v},t)$ of a test star, moving through a homogenous sea of field
stars. They are described by a static space-independent distribution
function, $\Psi({\bf v})$, which determines $D^{i}$ and $D^{ij}$. If
each field star have the same mass $m$ than the test star, the
Rosenbluth coefficients take the form \cite{BinTre,rosenbluth}
\begin{eqnarray}
    A^{i}_{N}&=&8\pi G^{2}m^{2}\ln \Lambda\frac{\partial}{\partial
   v_{i}}\int d^{3}{\bf v'}\frac{\Psi({\bf v'})}{|{\bf v}-{\bf
   v'}|},\label{rosen1}\\ &&\nonumber\\
   B^{ij}_{N}&=&4\pi G^{2}m^{2}\ln \Lambda\frac{\partial^{2}}{\partial
   v_{i}\partial
   v_{j}}\int d^{3}{\bf v'}\Psi({\bf v'})|{\bf v}-{\bf v'}|,\label{rosen2}
\end{eqnarray}
where $\ln \Lambda$ is the Coulomb logarithm and $G$ the gravitation
constant. As it is well known, derivation of
(\ref{rosen1})-(\ref{rosen2}) requires three fundamental facts: (i)
To assume that the binary weak encounters dominate the system's
relaxation process \cite{chan1,chan2}; (ii) to assume that all of
these encounters are local (i.e. with impact parameter much less
than the system size); (iii) a detailed knowledge of the two-body
microscopic dynamics.

Perhaps, a natural way to obtain an 1PN version of
(\ref{fokk-planck}), is to start from a covariant Fokker-Planck
equation and then perform its post-Newtonian approximation. At
present, it can be find different versions of this kind of kinteic
relations, such as the one derived by Kandrup
\cite{kand1,kand2,kand3} based on a stochastic standpoint (i.e. from
a covariant master equation). A more fundamental approach was
introduced recently by G. Chac\'on-Acosta and G.M. Kremer
\cite{chacon}, who derived a covariant Fokker-Planck equation for a
simple gas in the presence of gravitational field (see eq.
(\ref{fpcovariant1})), starting from the general-relativistic
Boltzmann equation. There are several reasons impelling us to adopt
this scheme. Since such derivation is supported on the Boltzmann
equation, the corresponding Fokker-Planck relation is consistent
with situations where the system evolves towards an equilibrium
sate, characterized by the Maxwell-J\"{u}tner distribution function
\cite{kremer}. On the other hand, such equation is valid for systems
whose relaxation process is dominated by \emph{grazing collisions},
which is the usual picture to model the behavior of self-gravitating
systems (in this context it is more convenient to use the term
\emph{encounters} instead of \emph{collisions}). More over, the
diffusion coefficients characterizing (\ref{fpcovariant1}) are
presented in a form that facilitates the implementation of the
post-Newtonian approach. We shall show this in section
\ref{covariant}, dedicated to obtain the 1PN approximation of
(\ref{fpcovariant1}). The result is an equation that is not
manifestly covariant (this is a characteristic of the 1PN scheme),
whose convective (l.h.s) and collision (r.h.s) terms can be split on
a Newtonian and post-Newtonian contributions.

The limitations of the above scheme can be viewed through the
consideration of a more fundamental standpoint, i.e. a BBGKY
formalism \cite{Bogoliubov,born-green,kirkwood,yvon}, which will be
the subject in later sections. By examining the system's dynamics at
a microscopic level, one can note that a fundamental aspect of the
1PN interaction has not been contemplated in the formulation of
(\ref{fpcovariant1}). As we shall show in section \ref{1PN-BBGKY
Equation}, the 1PN force exerted on each particle has a ternary
nature (see (\ref{mov1}),(\ref{gamaex})). This feature leads to the
apparition of 3-order correlation patterns in the first BBGKY
equation and not all of those vanish after the implementation of
binary encounters assumption (see (\ref{boltzmann}), (\ref{F})),
which is the basic statement employed in the formulation of
collision terms of the Landau \cite{landau,bal,Kunk} or
Fokker-Planck type.

Despite of the apparition of third order correlation patterns
originates a collision term that does not look like in a standard
fashion (eq. (\ref{F})), it can be put in a FPR form by choosing
adequately the two point correlation function $g_{2}$ and setting
$g_{3}=0$. Here, two facts will play an important role: (i) The
classical FPR equation can be derived from the first BBGKY equation
by choosing a correlation function corresponding to a weakly coupled
gas (WCG) in the hydrodynamical regime (it is not a surprising fact,
since in the usual derivation of the FPR equation it is assumed that
the weak encounters play a dominant role in the relaxation process);
(ii) in the 1PN approach the momentum conservation law for a system
of point particles is the same as in Newtonian theory, and it holds
if and only if each particle obeys the Newtonian equation of motion
\cite{Wein}. This means that the scattering process and, in
consequence, the explicit form of $g_{2}$ in 1PN approximation are
the same as in the Newtonian case. These considerations permit us to
incorporate the hydrodynamical WCG correlation function in the
post-Newtonian contribution of the collision term for its subsequent
simplification.

\section{Derivation from a Covariant Fokker-Planck Equation \label{covariant}}

We start by considering that the distribution function
$f(x^{\mu},p^{i})$ of a self-gravitating gas of particles with
identical rest mass $m$, satisfies  the covariant Fokker-Planck
equation \cite{chacon}:
\begin{equation}\label{fpcovariant1}
p^{\mu}\frac{\partial f}{\partial
x^{\mu}}-\Gamma^{\sigma}_{\mu\nu}p^{\mu}p^{\nu}\frac{\partial
f}{\partial p^{\sigma}}=-\frac{\partial }{\partial
p^{\mu}}(fD^{\mu})+\frac{1}{2}\frac{\partial^{2} }{\partial
p^{\mu}\partial p^{\nu}}(fD^{\mu\nu})
\end{equation}
where $p^{\mu}=mU^{\mu}$ is the four-momentum,
$\Gamma^{\sigma}_{\mu\nu}$ are the Christoffel's symbols,
$x^{\mu}=(ct, \mathbf{x})$ is the set of configuration coordinates
and $D^{\mu}$, $D^{\mu\nu}$ are the diffusion coefficients given by
\begin{eqnarray}
    D^{\mu}&=&\int f_{*}\Delta p_{*}^{\mu}
    F\sigma d\Omega \sqrt{-g}\frac{d^{3}p_{*}}{p_{*0}},\label{Acov1}\\ &&\nonumber\\
   D^{\mu\nu}&=&\int f_{*}\Delta p_{*}^{\mu}\Delta p_{*}^{\nu}
    F\sigma d\Omega \sqrt{-g}\frac{d^{3}p_{*}}{p_{*0}},\label{Dcov1}
\end{eqnarray}
where $f_{*}=f(x^{\mu},p_{*}^{i})$, $\Delta
p_{*}^{\mu}=p_{*}^{'\mu}-p_{*}^{\mu}$, $g=\det(g_{\mu\nu})$,
$F=\sqrt{(p_{*}^{\alpha}p_{\alpha})^{2}-m^{4}c^{4}}$ is the
invariant flux, $\sigma$ is the differential cross section and
$d\Omega$ is an element of solid angle which characterizes the
scattering process. The quantities $p^{\mu}$ and $p_{*}^{\mu}$
denote the four-momentum of the test and field stars before the
encounter, respectively, while $p^{'\mu}$ and $p_{*}^{'\mu}$
represent their four-momentum after the encounter.

In order to perform the 1PN approximation of (\ref{fpcovariant1}),
we have to take into account the following facts. We need an
expansion of (\ref{fpcovariant1})-(\ref{Dcov1}) up to order
$(\bar{v}/c)^{4}$, where $\bar{v}$ is a typical Newtonian speed in
the system and $c$ is the speed of light. In this approximation
$g_{\mu\nu}$ is given in terms of the Newtonian field $\Phi$ and the
post-newtonian fields $\psi$, $\xi_{i}$ (see \ref{ap1}) :
\begin{eqnarray}
  g_{00} & = & -1-2\Phi/c^{2}-2(\Phi^{2}+\psi)/c^{4},\label{g00}\\
  g_{0i} & = & \xi_{i}/c^{3}, \label{gi0}\\
  g_{ij} & = & (1-2\Phi/c^{2})\delta_{ij}\label{gij}.
\end{eqnarray}
(Latin indices run from 1 to 3). On the other hand, the
four-velocity $U^{\mu}$ is related to the classical velocity
$v^{i}=dx^{i}/dt$ through the equation
\begin{equation}\label{velocityvector}
U^{\mu}=U^{0}V^{\mu},\:\:\:\:\:\:\:\:V^{\mu}=(1,v^{i}/c),
\end{equation}
and it is restricted by the relation
\begin{equation}\label{restriction}
g_{\mu\nu}U^{\mu}U^{\nu}=-c^{2}.
\end{equation}
Another important fact is that in the 1PN approximation the momentum
conservation for a system of point particles is satisfied if and
only if each particle obeys the Newtonian equation of motion
\cite{Wein}. In other words: they obey the classical momentum
conservation law. Immediately we note two implications that will
play a fundamental role in the calculation of the diffusion
coefficients. They were obtained in reference \cite{chacon} by
choosing the center-of-mass system corresponding to two colliding
particles. Since in 1PN approximation they obey the classical
momentum conservation law, it implies that $\Delta U^{0}_{*}=0$ and,
in consequence, $D^{0}=D^{00}=0$. On the other hand, also in virtue
of the classical momentum conservation law, we must to consider
$\sigma d\Omega$ as the usual Newtonian scattering cross section.
The above considerations enable us to rewrite
(\ref{fpcovariant1})-(\ref{Dcov1}) as
\begin{equation}\label{fpcovariant2}
\mathcal{L}_{U}f=-\frac{\partial }{\partial
U^{i}}(fA^{i})+\frac{1}{2}\frac{\partial^{2} }{\partial
U^{i}\partial U^{j}}(fB^{ij})
\end{equation}
where $\mathcal{L}_{U}$ is the Liouville's operator defined as
\begin{equation}\label{operadorL}
\mathcal{L}_{U}=U^{\mu}\frac{\partial }{\partial
x^{\mu}}-\Gamma^{i}_{\mu\nu}U^{\mu}U^{\nu}\frac{\partial }{\partial
U^{i}}
\end{equation}
and
\begin{eqnarray}
    A^{i}&=&\int f_{*}\Delta U_{*}^{i}
    \sqrt{(U_{*}^{\alpha}U_{\alpha})^{2}-c^{4}}\tilde{\sigma} d\Omega \sqrt{-g}\frac{d^{3}U_{*}}{U_{*0}},\label{Acov2}\\ &&\nonumber\\
   B^{ij}&=&\int f_{*}\Delta U_{*}^{i}\Delta U_{*}^{j}
    \sqrt{(U_{*}^{\alpha}U_{\alpha})^{2}-c^{4}}\tilde{\sigma} d\Omega \sqrt{-g}\frac{d^{3}U_{*}}{U_{*0}}.\label{Dcov2}
\end{eqnarray}
We have decided to express the Fokker-Planck equation in terms of
the four-velocity instead of the four-momentum in order to
facilitate the implementation of some results used by Rezania and
Sobouti in the collisionless case \cite{rezania}. Now
$f(x^{\mu},U^{i})$ is a phase density in the six dimensional
$(x^{i}, U^{i})$-space and $\tilde{\sigma}$ is the Newtonian
differential cross section. The l.h.s. of (\ref{fpcovariant2}) in
1PN approximation can be written as \cite{rezania}
\begin{eqnarray}
 \mathcal{L}_{U}f&=& \frac{U^{0}}{c}\left\{\frac{\partial f}{\partial t} +v^{i}\frac{\partial f}{\partial x^{i}}
  -\frac{\partial \Phi}{\partial x^{i}}\frac{\partial f}{\partial v^{i}}
  -\frac{1}{c^{2}}\left[(4\Phi+v^{2})\frac{\partial \Phi}{\partial x^{i}}
  \right.\right.\nonumber\\
   &&\nonumber\\
  &&\left.\left.-v^{i}v^{j}\frac{\partial \Phi}{\partial x^{j}}
  -v^{i}\frac{\partial \Phi}{\partial t}+\frac{\partial \psi}{\partial x^{i}}
  +\left(\frac{\partial \xi_{i}}{\partial x^{j}}-\frac{\partial \xi_{j}}{\partial x^{i}}\right)v^{j}
  +\frac{\partial \xi_{i}}{\partial t}\right]\frac{\partial f}{\partial v^{i}}
  \right\},
\end{eqnarray}
where $U^{0}/c$, determined from (\ref{restriction}), is given by
\begin{equation}\label{Uo}
\frac{U^{0}}{c}=1+\frac{v^{2}}{2c^{2}}-\frac{\Phi}{c^{2}}.
\end{equation}
In consequence, equation (\ref{fpcovariant2}) can be put in the form
\begin{equation}\label{fpcovariant3}
(\mathcal{L}_{N}+\mathcal{L}_{PN})f=\lambda\left[-\frac{\partial
}{\partial U^{i}}(fA^{i})+\frac{1}{2}\frac{\partial^{2} }{\partial
U^{i}\partial U^{j}}(fB^{ij})\right]
\end{equation}
where $\mathcal{L}_{N}$ and $\mathcal{L}_{PN}$ are the classical and
post-Newtonian Liouville's operators, given by
\begin{eqnarray}
 \mathcal{L}_{N}&=& \frac{\partial }{\partial t} +v^{i}\frac{\partial }{\partial x^{i}}
  -\frac{\partial \Phi}{\partial x^{i}}\frac{\partial }{\partial
  v^{i}}\label{operadorLc}
  \\
   &&\nonumber\\
  \mathcal{L}_{PN}&=&-\frac{1}{c^{2}}\left[(4\Phi+v^{2})\frac{\partial \Phi}{\partial x^{i}}
  -v^{i}v^{j}\frac{\partial \Phi}{\partial x^{j}}
  -v^{i}\frac{\partial \Phi}{\partial
  t}\right.\nonumber\\&&\nonumber\\
  &&\left.\:\:\:\:\:\:\:\:\:\:\:\:\:\:\:+\frac{\partial \psi}{\partial x^{i}}
  +\left(\frac{\partial \xi_{i}}{\partial x^{j}}-\frac{\partial \xi_{j}}{\partial x^{i}}\right)v^{j}
  +\frac{\partial \xi_{i}}{\partial t}\right]\frac{\partial }{\partial v^{i}}.\label{operadorLpn}
\end{eqnarray}
and $\lambda=c/U^{0}$, up to order $c^{-2}$, equals to
\begin{equation}\label{lamda}
\lambda=1-\frac{v^{2}}{2c^{2}}+\frac{\Phi}{c^{2}}.
\end{equation}
According to (\ref{fpcovariant3}), we need an expansion of the
Fokker-Planck operator (the term in parenthesis in the r.h.s.) up to
order $c^{-2}$. Here we have to take into account that $\partial
v^{i}/\partial U^{j}$ is given by \cite{rezania}
\[\frac{\partial v^{j}}{\partial U^{i}}= \:\: \left\{ \begin{array}{ll}
    Q^{-1}v^{j}(g_{0i}+g_{ik}v^{k}/c)
    \:\:\:\:\:\:\:\:\:\:\:\:\:\:\:\:\:\:\:\:\:\:\:\:\:\:\:\:\:\:\:\:\:\:\:\:\:\:\:\:\mbox{for}\:\:i\neq k
    \\ \\ -Q^{-1}[c^{3}U^{0^{-2}}+\sum_{k\neq i}v^{k}(g_{0k}+g_{lk}v^{l}/c)]  \:\:\:\:\:\:\:\:\mbox{for}\:\:i= k \end{array}\right.
\]
where
\begin{equation}\label{Qtrans}
Q=U^{0}(g_{00}+g_{0l}v^{l}/c).
\end{equation}
Then we obtain
\begin{equation}\label{operator1}
\frac{\partial}{\partial U^{i}}= \frac{\partial v^{j}}{\partial
U^{i}}\frac{\partial}{\partial v^{j}}=\beta\frac{\partial}{\partial
v^{i}}+\frac{1}{c^{2}}\left(\sum_{k\neq
i}v^{k}v^{k}\frac{\partial}{\partial v^{i}}-\sum_{j\neq
i}v^{i}v^{j}\frac{\partial}{\partial v^{j}}\right)
\end{equation}
with
\begin{eqnarray}
  \beta & = & 1-\frac{3v^{2}}{2c^{2}}+
   \frac{\Phi}{c^{2}}.
\end{eqnarray}

Now we shall obtain $A^{i}$ and $B^{ij}$ up to order $c^{-2}$. It is
useful to consider that in the center-of-mass system
$U_{*}^{0}=U^{0}$, $U_{*}^{i}=-U^{i}$ and, as a consequence of the
condition (\ref{restriction}), we have that $g_{00}=g_{00*}$,
$g_{0i}=-g_{0i*}$, $g_{ij}=g_{ij*}$. This will help us to simplify
the calculation of quantities like $\Delta U_{*}^{i}/U_{0*}$,
$d^{3}U_{*}$ and $U_{*}^{\alpha}U_{\alpha}$. In fact, after some
calculations, we find
$$
\frac{\triangle
U_{*}^{i}}{U_{*0}}=\left(-1+\frac{2\Phi}{c^{2}}\right)\frac{\triangle
v^{i}_{*}}{c},\quad\sqrt{-g}=1-\frac{2\Phi}{c^{2}},\quad
d^{3}U_{*}=\left(1+\frac{5v^{2}}{2c^{2}}-\frac{3\Phi}{c^{2}}\right)d^{3}v_{*},
$$
and
$$
\sqrt{(U_{*}^{\alpha}U_{\alpha})^{2}-c^{4}}=
c\left(1+\frac{v^{2}}{2c^{2}}-\frac{4\Phi}{c^{2}}\right)|\mathbf{v}-\mathbf{v}_{*}|
$$
By introducing the above relations in (\ref{Acov2})-(\ref{Dcov2}),
we obtain up to order $c^{-2}$
\begin{eqnarray}
A^{i}&=&\int\gamma|\mathbf{v}-\mathbf{v}_{*}| f_{*}\Delta
v_{*}^{i}\tilde{\sigma} d\Omega
d^{3}v_{*},\label{Ai1PN}\\
B^{ij}&=&\int\eta|\mathbf{v}-\mathbf{v}_{*}| f_{*}\Delta
v_{*}^{i}\Delta v_{*}^{j}\tilde{\sigma} d\Omega
d^{3}v_{*},\label{Bij1PN}
\end{eqnarray}
with
\begin{equation}\label{gamaint}
\gamma=-1-\frac{7v^{2}}{2c^{2}}+\frac{11\Phi}{c^{2}},\quad\quad\eta=-1-\frac{4v^{2}}{c^{2}}+\frac{12\Phi}{c^{2}}.
\end{equation}
From equations (\ref{Ai1PN})-(\ref{Bij1PN}) one can see the relation
between $A^{i}$, $B^{ij}$ and the classical Rosenbluth coefficients.
In order to do this one has to assume that $f_{*}$ can be replaced
by a distribution function of a homogeneous field stars population
in equilibrium and to consider the differential cross section
corresponding to the gravitational inverse-square law of force. That
is
\begin{equation}\label{fysigma}
f(\mathbf{x},\mathbf{v}_{*},t)\rightarrow
\Psi(\mathbf{v}_{*}),\quad\quad\tilde{\sigma}=G^{2}m^{2}|\mathbf{v}-\mathbf{v}_{*}|^{-4}\sin^{-4}(\theta/2),
\end{equation}
where $\Psi(\mathbf{v}_{*})=\Psi_{*}$ is the field star DF and
$\theta$ is the scattering angle as measured in the center-of-mass
frame. By implementing (\ref{fysigma}) in
(\ref{Ai1PN})-(\ref{Bij1PN}) one can see that the $c^{0}$-factors of
$A^{i}$ and $B^{ij}$ are equivalent to the Rosenbluth's coefficients
$A^{i}_{N}$ and $B^{ij}_{N}$, given by (\ref{rosen1})-(\ref{rosen2})
(details of this calculation can be seen in \cite{rosenbluth}).
Therefore, we can write
\begin{equation}\label{AyB}
A^{i}=-A^{i}_{N}-A^{i}_{PN},\quad\quad
B^{ij}=-B^{ij}_{N}-B^{ij}_{PN},
\end{equation}
where $A^{i}_{N}$ and $B^{ij}_{N}$ are given by
(\ref{rosen1})-(\ref{rosen2}) and
\begin{eqnarray}
A^{i}_{PN}&=&\frac{G^{2}m^{2}}{c^{2}}\int\left(\frac{7v^{2}}{2c^{2}}-\frac{11\Phi}{c^{2}}\right)
  \frac{\Psi(\mathbf{v}_{*})\Delta
v_{*}^{i}d\Omega
d^{3}v_{*}}{|\mathbf{v}-\mathbf{v}_{*}|^{3}\sin^{4}(\theta/2)}
,\label{Ai2PN}\\&&\nonumber\\
 B^{ij}_{PN}&=&\frac{G^{2}m^{2}}{c^{2}}
\int\left(\frac{4v^{2}}{c^{2}}-\frac{12\Phi}{c^{2}}\right)\frac{\Psi(\mathbf{v}_{*})\Delta
v_{*}^{i}\Delta v_{*}^{j}d\Omega
d^{3}v_{*}}{|\mathbf{v}-\mathbf{v}_{*}|^{3}\sin^{4}(\theta/2)}.\label{Bij2PN}
\end{eqnarray}
Finally, introducing (\ref{AyB}) and (\ref{operator1}) in
(\ref{fpcovariant3}), we find that the 1PN approximation of
(\ref{fpcovariant2}) can be written as
\begin{eqnarray}
  (\mathcal{L}_{N}+\mathcal{L}_{PN})f & = & -\frac{\partial}{\partial
   v_{i}}(A^{i}_{N}f)+\frac{1}{2}\frac{\partial^{2}}{\partial v_{i}\partial
   v_{j}}(B^{ij}_{N}f) \nonumber\\&&\nonumber\\
   &  &-\frac{\partial}{\partial
   v_{i}}(A^{i}_{PN}f)+\frac{1}{2}\frac{\partial^{2}}{\partial v_{i}\partial
   v_{j}}(B^{ij}_{PN}f)
   \nonumber\\&&\nonumber\\
   &
   &+\frac{1}{c^{2}}\left[\mathcal{J}_{i}(A^{i}_{N}f)+\mathcal{K}_{ij}(B^{ij}_{N}f)\right],\label{fp1PN}
\end{eqnarray}
where we have defined the operators $\mathcal{J}_{i}$ and
$\mathcal{K}_{ij}$ as
\begin{eqnarray}
  \mathcal{J}_{i}& = & (v^{2}-2\Phi)\frac{\partial}{\partial
   v^{i}}+\delta_{ij}v^{j}\left(2+v^{l}\frac{\partial}{\partial v^{l}}\right), \label{opJ}\\&&\nonumber\\
  \mathcal{K}_{ij}&=&\left(v^{2}-\frac{11\Phi}{4}\right)\frac{\partial^{2}}{\partial v_{i}\partial
   v_{j}}+\frac{3v^{i}v^{l}}{4}\frac{\partial^{2}}{\partial v_{l}\partial
   v_{j}}
   \nonumber\\&&\nonumber\\
   &  &+\delta_{ij}\left(2+v^{l}\frac{\partial}{\partial
   v^{l}}\right)+\frac{11v^{i}}{2}\frac{\partial}{\partial
   v^{j}}.\label{opK}
\end{eqnarray}
The relation (\ref{fp1PN}), in contrast with (\ref{fpcovariant1}),
is not a manifestly covariant kinetic equation that can be
interpreted as an 1PN extension of the classical FPR equation
(\ref{fokk-planck}). There, the Newtonian and post-Newtonian
contributions appear separated both in the l.h.s and in the r.h.s.
We note that the post-Newtonian fields $\psi$ and $\xi_{i}$
contribute only through $\mathcal{L}_{PN}$ and do not appear in the
collision term. In the next sections we will see that a derivation of
the kinetic equation from microscopic dynamics reveals additional contributions
to the diffusion coefficients. The reason is that the 1PN interaction has ternary
contributions, coming from $\psi$ and $\xi_{i}$, that do not disappear when the
BBGKY sequence is closed.

\section{Derivation from the Microscopic Dynamics:
The First 1PN-BBGKY Equation \label{1PN-BBGKY Equation}}

In this section, we will derive the statistical picture of the
self-gravitating system starting from its microscopic dynamics. In
particular, we shall deal with the BBGKY hierarchy that formally
equals to the Liouville's equation. In particular, we shall consider
only the first equation of the sequence, from which we will obtain
the kinetic equation. Since we start from the non-covariant 1PN law
of force, the corresponding evolution equation will be also
expressed in a non-manifestly covariant fashion. As it was pointed
in the previous section, this is a feature characterizing the 1PN
approximation.

Let a system composed by $N$ identical point particles, with mass
$m$, interacting gravitationally and moving with velocity $\ll c$.
According to the 1PN approximation, each particle experiments an
acceleration given by
\begin{equation}\label{mov1}
\dot{{\bf v}} = {\bf g}_{K}({\bf x},t) + {\bf \Gamma}_{K}({\bf
x},{\bf v},t),
\end{equation}
where ${\bf g}_{K}$ and ${\bf \Gamma}_{K}$  are the newtonian and
post Newtonian gravitational forces (per unit mass), respectively
(the subscript $K$, motivated by the subsequent use of Klimontovich
functions, indicates that we deal with exact fields instead of mean
fields). They are given by \cite{Wein}
\begin{equation}\label{gkli}
{\bf g}_{K} =
 -Gm\sum_{i=1}^{N}\frac{{\bf x}-{\bf x}_{i}}{|{\bf x}-{\bf x}_{i}|},
\end{equation}
\begin{eqnarray}
c^{2} {\bf \Gamma}_{K} &=& - \frac{\partial}{\partial {\bf x}}(2\Phi
^{2} + \psi ) - \frac{\partial {\cal\xi}}{\partial t} + {\bf v}
\times \left(\frac{\partial}{\partial {\bf x}} \times {\cal\xi}
\right)
\nonumber \\
&& \label{gama} \\
&&+ \ 3 {\bf v} \frac{\partial \Phi}{\partial t} + 4 {\bf v} \left(
{\bf v} \cdot \frac{\partial}{\partial {\bf x}} \right) \Phi - {\bf
v}^{2} \frac{\partial \Phi}{\partial {\bf x}}. \nonumber
\end{eqnarray}
The last equation represents the relativistic correction to the
force in the 1PN approximation, which includes the contribution of
the post-Newtonian potentials ${\cal\xi}$ and $\psi$ . In this case
(identical point-like masses), we can obtain an explicit form for
${\bf \Gamma}_{K}$, in terms of positions and velocities. We have
found that this post Newtonian contribution can be written in a very
suggestive form (see \ref{ap1}):
\begin{equation}\label{gamaex}
{\bf \Gamma}_{K}= \sum_{i=1}^{N} {\bf \Lambda}({\bf w},{\bf w}_{i})
+ \sum_{i=1}^{N} \sum_{j\neq i} {\bf \Upsilon}({\bf x},{\bf
x}_{i},{\bf x}_{j}),
\end{equation}
where ${\bf w}\equiv ({\bf x},{\bf v})$ (this notation will be used
henceforward). Here, the total relativistic force is shown as the
result of two contributions: a velocity-dependent binary interaction
term ${\bf \Lambda}$, and a velocity-independent ternary interaction
term ${\bf \Upsilon}$. They are defined as follows
\begin{eqnarray}
{\bf \Lambda}({\bf w},{\bf w}_{i}) &\equiv& \frac{Gm}{c^{2}} \left\{
\frac{{\bf r}_{xx_{i}}}{r_{xx_{i}}^{3}} \left[\frac{4 m
G}{r_{xx_{i}}} - {\bf v}^{2} - 2
{\bf v}_{i}^{2} + 4 {\bf v} \cdot {\bf v}_{i} \right. \right. \nonumber \\
&& \nonumber \\
&&\left. +\frac{3}{2} \left(\frac{{\bf v}_{i} \cdot {\bf
r}_{xx_{i}}}{r_{xx_{i}}}\right)^{2}  \right]\left. + ({\bf v} - {\bf
v}_{i}) \frac{(4 {\bf v} - 3 {\bf v}_{i}) \cdot
{\bf r}_{xx_{i}}}{r_{xx_{i}}^{3}} \right\}, \label{Lambda}   \\
&& \nonumber \\
&&  \nonumber   \\
{\bf \Upsilon}({\bf x},{\bf x}_{i},{\bf x}_{j}) &\equiv& \frac{G^{2}
m^{2}}{2c^{2}r_{xx_{i}}} \left[ 8\frac{{\bf
r}_{xx_{j}}}{r_{xx_{j}}^{3}} - 7\frac{{\bf
r}_{x_{i}x_{j}}}{r_{x_{i}x_{j}}^{3}} \right.\left. + {\bf
r}_{xx_{i}} \frac{2r_{x_{i}x_{j}}^{2}-{\bf r}_{xx_{i}}\cdot{\bf
r}_{x_{i}x_{j}}}{r_{xx_{i}}^{2}r_{x_{i}x_{j}}^{3}}
\right],\label{Upsilon}
\end{eqnarray}
with
\begin{equation}\label{r}
    {\bf r}_{xx_{i}}={\bf x}-{\bf x}_{i}.
\end{equation}
As we shall show, ${\bf \Lambda}$ leads to the incorporation of the
two-point correlation function in the statistical description. On
the other hand, ${\bf \Upsilon}$ causes the apparition of third
order correlation patterns, which is an essential difference  with
the purely Newtonian case.

In order to introduce the statistical description of the evolution
for this system, we use the KD approach \cite{Klim,Dup}. Such
formulation of non-equilibrium statistical mechanics, had been
widely used in the probabilistic treatment of systems dominated by
interactions of kind (\ref{mov1}) (see \cite{zim,golds,MaBer}, as
examples). The method starts by introducing the microscopic one
particle phase space density (KD function):
\begin{equation}\label{kli}
f_{K}({\bf x},{\bf v},t) = \sum_{i=1}^{N} \delta[{\bf x}-{\bf
x}_{i}(t)] \delta[{\bf v} - {\bf v}_{i}(t)],
\end{equation}
where $\delta$ is the 3-dimensional Dirac delta function. Clearly,
this function satisfies the following evolution equation (KD
equation):
\begin{equation}\label{eckli}
\frac{\partial f_{K}}{\partial t} + {\bf v} \cdot \frac{\partial
f_{K}}{\partial {\bf x}} + {\bf g}_{K} \cdot \frac{\partial
f_{K}}{\partial {\bf v}} + {\bf \Gamma}_{K}\cdot\frac{\partial
f_{K}}{\partial {\bf v}}= 0,
\end{equation}
where $g_{K}$ and ${\bf \Gamma}_{K}$ are given by (\ref{gkli}) and
(\ref{gamaex}), or by the equivalent relations
\begin{eqnarray}
{\bf g}_{K}({\bf x},t) &=& - Gm \int d^{6}{\bf w'} f_{K}({\bf
w'},t) \frac{{\bf r}_{xx'}}{ r_{xx'}^{3}}, \label{gkli2} \\
&&  \nonumber   \\
{\bf \Gamma}_{K}({\bf w},t) &=& \int d^{6}{\bf w'} f_{K}({\bf
w'},t) {\bf \Lambda}({\bf w},{\bf w'}) \nonumber \\
&&   \nonumber \\
&&+\int d^{6}{\bf w'} d^{6}{\bf w''} f_{K} ({\bf w'},t) f_{K} ({\bf
w''},t){\bf \Upsilon}({\bf x},{\bf x'},{\bf x''}). \label{gamakli}
\end{eqnarray}

The one particle distribution function (representing the system
state), as much as the first BBGKY equation (describing its temporal
evolution), are obtained averaging the KD function and the KD
equation, respectively. Indeed, it is easy to see that the average
of $f_{K}$ is
\begin{equation}\label{relfklif}
\langle f_{K}({\bf w},t) \rangle = f({\bf w},t),
\end{equation}
where $f$ is the usual one particle distribution function, which
represents the behavior of an average particle in the system.  Also,
there exist relations connecting
 correlation functions, $f$ and $f_{K}$ \cite{bal,golds}:
\begin{equation}\label{prom2f2}
\langle f_{K} f'_{K} \rangle = f f'+ g_{2}({\bf w},{\bf w'},t) +
\delta({\bf w'} - {\bf w}) f ,
\end{equation}
\begin{eqnarray}
\langle f_{K} f'_{K} f''_{K} \rangle &=& f f' f'' + g_{3}({\bf
w},{\bf w'},{\bf w''},t)
+f g_{2}({\bf w'},{\bf w''},t)+f' g_{2}({\bf w},{\bf w''},t) \nonumber\\
&&  \nonumber   \\
&&  +f'' g_{2}({\bf w},{\bf  w'},t)+\delta({\bf w}-{\bf w'})[ff''+g_{2}({\bf w},{\bf w''},t)]\nonumber \\
&&  \nonumber   \\
&& +\delta({\bf w}-{\bf w''})[ff'+g_{2}({\bf w},{\bf w'},t)]+\delta({\bf w}-{\bf w'})\delta({\bf w}-{\bf w''})f \nonumber \\
&&  \nonumber   \\
&& +\delta({\bf w'}-{\bf w''})[ff'+g_{2}({\bf w},{\bf
w'},t)]\label{prom2f3}.
\end{eqnarray}
Here $g_{2}$ and $g_{3}$ are the two and three point correlation
functions, respectively, and $f=f({\bf w},t)$, $f'=f({\bf w'},t)$,
etc.

As it is usual, we assume that $f$ must satisfy a kinetic equation
describing its evolution. In order to derive such equation, we start
from the BBGKY first equation, which is obtained by taking the
average of (\ref{eckli}), using  relations
(\ref{relfklif})-(\ref{prom2f3}):
\begin{equation}\label{boltzmann}
\frac{\partial f}{\partial t} + {\bf v} \cdot \frac{\partial
f}{\partial {\bf x}} + {\bf g} \cdot \frac{\partial f}{\partial {\bf
v}} + {\bf \Gamma}\cdot \frac{\partial f}{\partial {\bf v}}  = -
\frac{\partial }{\partial {\bf v}} \cdot {\bf F},
\end{equation}
where ${\bf g}$, ${\bf \Gamma}$ (newtonian and post Newtonian
average gravitational fields) are given by
\begin{eqnarray}
{\bf g}({\bf x},t) &=& - Gm \int d^{6}{\bf w'} f' \frac{{\bf
r}_{xx'}}{ r_{xx'}^{3}},\label{g}\\&&\nonumber\\ {\bf \Gamma}({\bf
w},t) &=& \int d^{6}{\bf w'} f' {\bf \Lambda}({\bf w},{\bf w'})+
\int d^{6}{\bf w'} d^{6}{\bf w''} f' f'' {\bf \Upsilon}({\bf x},{\bf
x'},{\bf x''}), \label{gama2}
\end{eqnarray}
while in the r.h.s. (collision term), ${\bf F}$ has the form
\begin{eqnarray}\label{F}
{\bf F} &\equiv & \int d^{6}{\bf w'} g_{2}({\bf w},{\bf w'},t)
\left[{\bf \Lambda}({\bf w},{\bf w'}) - Gm\frac{{\bf r}_{xx'}}{
r_{xx'}^{3}}
\right]  \\
&&  \nonumber   \\
& & +\int d^{6}{\bf w'} d^{6}{\bf w''} \left[ g_{2}({\bf w'},{\bf
w''},t)
f + g_{2}({\bf w},{\bf w''},t) f' \right. \nonumber \\
&&  \nonumber   \\
& & \left. +g_{2}({\bf w},{\bf w'},t) f'' +g_{3}({\bf w},{\bf
w'},{\bf w''},t)\right]
  {\bf \Upsilon}({\bf x},{\bf x'},{\bf x''}).\nonumber
\end{eqnarray}
Equation (\ref{boltzmann}) is the first equation of the 1PN-BBGKY
hierarchy. The l.h.s., (convective term) depends on the mean fields
${\bf g}$, ${\bf \Gamma}$, and has the same structure of a mean
field kinetic term (the l.h.s. of a Vlasov equation
\cite{vlasov,golse}). More over, it is easy to see that this
convective term is equivalent to the one characterizing eq.
(\ref{fpcovariant3}). However, the collision term is given in a
non-standard form. We can see it as a sum of two contributions: (i)
a two-point term, characterized by $g_{2}$ and binary interactions
${\bf g}$, ${\bf \Lambda}$; (ii) a three-point term, characterized
by the ternary interaction ${\bf \Upsilon}$ and ternary correlation
patterns of the form $fg_{2}$ and $g_{3}$. This last contribution
makes non trivial the task of close the hierarchy, in order to
derive a tractable kinetic equation. However, as it will be shown
later, we find that is possible to do this, establishing some
reasonable assumptions about the correlation functions behavior. In
particular, we will assume that the star cluster relaxation process
follows a sort of WCG correlation dynamics. Such assumption comes
from the fact that the classical FPR equation can be derived
directly from the first BBGKY equation by introducing a two point
correlation function corresponding to a WCG in hydrodynamical
regime. The validity of such statement will be the subject of the
next section.

\subsection{An Alternative Derivation of the Classical FPR equation:
the Hydrodynamical-WCG Approximation  \label{collterm}}

As it was pointed out in sect.\ref{intro}, weak encounters play a
dominant role in the star cluster relaxation process
\cite{chan1,chan2}. This fact suggests that we can expect a close
analogy between the star cluster relaxation process (dominated by
weak encounters) and the relaxation process of a WCG (based on weak
interactions). In fact, we can show that the introduction of a WCG
correlation function (in the hydrodynamic regime) in the collision
term of the classical first BBGKY relation, leads to the usual FPR
equation. Before to do this, we expose briefly some relevant aspects
of the WCG correlation dynamics.

A classical WCG is commonly described as a set of particles
interacting via gaussian potentials of the form \cite{bal}
\begin{equation}\label{Vwcg}
    V(r)=V_{o}\exp[-(r/\alpha)^{2}],
\end{equation}
where $r$ is the inter-particle separation, $V_{o}$ represents the
potential's maximum value and $\alpha$ is an \emph{interaction
length} (for $r$ greater than $\alpha$ the interaction practically
vanishes).  The evolution in time, for this class of systems, is
described by \emph{the proper kinetic equation to order two} (see
\cite{bal}, p. 574) and there is an aspect we are specially
interested, around such equation: the explicit form of correlation
functions. Here we note that (to order two) $g_{3}$, $g_{4}$, etc,
vanish and that $g_{2}$ is given by the relation \cite{bal}:
\begin{equation}\label{g2wcg}
    g_{2}({\bf w},{\bf w'},t)= \int_{0}^{\infty}d\tau
    \frac{\partial V({\bf w},{\bf w'},\tau)}{\partial{\bf x}}\cdot\left(\nabla_{vv'}
    +\tau\nabla_{xx'}\right)[ff'],
\end{equation}
where we have used the notation
\begin{equation}\label{nablacorr}
    \nabla_{xx'}\equiv\frac{\partial}{\partial{\bf
    x}}-\frac{\partial}{\partial{\bf x'}},\:\:\:\:\:\:\:\:\:\:\:
    \nabla_{vv'}\equiv\frac{\partial}{\partial{\bf
    v}}-\frac{\partial}{\partial{\bf v'}},\:\:\:\:\:\:\:\:\:\:\:
    {\bf u}_{vv'}\equiv{\bf v}-{\bf v'}.
\end{equation}
and
\begin{equation}\label{potV}
    V({\bf w},{\bf w'},\tau)=V_{o}e^{-(|{\bf r}_{xx'}-\tau{\bf
    u}_{vv'}|/\alpha)^{2}}.
\end{equation}
Moreover, if we consider that the WCG is in the hydrodynamic regime,
then the spatial variation of $f$ is characterized by a
hydrodynamical length $L_{h}$, defined as \cite{bal}
\begin{equation}\label{Lh}
    L_{h}=\mbox{max}\frac{f}{|\partial f/\partial{\bf x}|}\gg
    \alpha.
\end{equation}
Since $L_{h}\gg\alpha$, the system can be considered practically
homogeneous ($f$ is spatially independent), over a distance of the
order of $\alpha$. We may take advantage of this fact in order to
simplify (\ref{g2wcg}). In the Taylor expansion of $f'$ at ${\bf
x'}$ around ${\bf x}$,
\begin{equation}\label{taylor}
    f({\bf x'},{\bf v'},t)=f({\bf x},{\bf v'},t)-{\bf r}_{xx'}\cdot\frac{\partial f({\bf x},{\bf v'},t)}{\partial{\bf
    x}}+\cdots,
\end{equation}
the second term is of the order $\alpha/L_{h}$ compared to the first
(the next terms are smaller). So we can take $f({\bf x'},{\bf
v'},t)\approx f({\bf x},{\bf v'},t)$ and then
$\partial[ff']/\partial{\bf x'}\approx 0$. Performing a similar
procedure, starting from the expansion of $f$ at ${\bf x}$ around
${\bf x'}$, we also obtain $\partial[ff']/\partial{\bf x}\approx 0$.
This means that, in the hydrodynamical limit, we can set
$\nabla_{xx'}[ff']\approx 0$ and replace $f'$ by $f({\bf x},{\bf
v'},t)$ in the remaining factor $\nabla_{vv'}[ff']$ of eq.
(\ref{g2wcg}). Thus, by carrying out the integral (\ref{g2wcg}),
$g_{2}$ can be written as
\begin{equation}\label{g2wcg2}
    g_{2}({\bf w},{\bf w'},t)= \mathbf{\mathcal{G}}({\bf w},{\bf w'})\cdot\nabla_{vv'}
    [f({\bf x},{\bf v'},t)f({\bf x},{\bf v},t)],
\end{equation}
with
\begin{eqnarray}\label{Gcal}
  \mathbf{\mathcal{G}}({\bf w},{\bf w'})& \equiv & \frac{\sqrt{\pi}V_{o}}{u_{vv'}^{2}}e^{-(r_{xx'}/\alpha)^{2}}
  \left\{\frac{{\bf u}_{vv'}}{\sqrt{\pi}}+(Q_{ww'}{\bf u}_{vv'}\right.\nonumber\\
   &  & \nonumber\\
   & & \left.-u_{vv'}{\bf r}_{xx'}/\alpha)\left[1+\mbox{Erf}(Q_{ww'})\right] e^{Q_{ww'}^{2}}
   \right\},
\end{eqnarray}
where $\mbox{Erf}(Q_{ww'})$ is the error function and
\begin{equation}\label{Qww}
    Q_{ww'}=\frac{{\bf u}_{vv'}\cdot{\bf r}_{xx'}}{u_{vv'}\alpha}.
\end{equation}

Now let us contemplate the implications concerning with the
implementation of (\ref{g2wcg2}) in the collision term of the
classical first BBGKY equation, which we write as
\begin{equation}\label{FN1}
   -\frac{\partial}{\partial{\bf v}}\cdot{\bf F}_{N}=- Gm\int d^{6}{\bf w'} g_{2}({\bf w},{\bf w'},t)
\frac{{\bf r}_{xx'}}{ r_{xx'}^{3}}.
\end{equation}
Again the subscript $N$ denotes that we are dealing with the
Newtonian case. In \ref{ap2} we show that after the introduction of
(\ref{g2wcg2}) in the above equation, it reduces to
\begin{equation}\label{FN2}
   -\frac{\partial}{\partial{\bf v}}\cdot{\bf F}_{N}=-\frac{\partial}{\partial
   v_{i}}(\tilde{A}^{i}_{N}f)+\frac{1}{2}\frac{\partial^{2}}{\partial v_{i}\partial
   v_{j}}(\tilde{B}^{ij}_{N}f)
\end{equation}
where we have introduced the definitions
\begin{eqnarray}
    \tilde{A}^{i}_{N}&=&-4\pi^{3/2}\alpha V_{o}Gm\frac{\partial}{\partial
   v_{i}}\int d^{3}{\bf v'}\frac{f({\bf x},{\bf v'},t)}{|{\bf v}-{\bf
   v'}|},\label{A2}\\ &&\nonumber\\
   \tilde{B}^{ij}_{N}&=&-2\pi^{3/2}\alpha V_{o}Gm\frac{\partial^{2}}{\partial
   v_{i}\partial
   v_{j}}\int d^{3}{\bf v'}f({\bf x},{\bf v'},t)|{\bf v}-{\bf v'}|.\label{B2}
\end{eqnarray}
From the above relations, we immediately  note that (\ref{FN2})
equals to the classical FPR collision term (\ref{fokk-planck}) by
choosing $V_{o}$ as
\begin{equation}\label{beta}
 V_{o}=-\frac{2Gm}{\alpha\sqrt{\pi}}\ln \Lambda,
\end{equation}
and assume that
\begin{equation}\label{fcampo}
 f({\bf x},{\bf v'},t)\approx\Psi({\bf v'}),
\end{equation}
where $\Psi({\bf v'})$, in the context of the FPR equation derived
in \S 8.1 of \cite{BinTre}, is a field star distribution function.
With (\ref{beta})-(\ref{fcampo}), $A^{i}_{N}$ and $B^{ij}_{N}$ are
exactly the Rosenbluth diffusion coefficients
(\ref{rosen1})-(\ref{rosen2}), describing the drift and diffusivity
generated by a homogeneous sea of background particles of mass $m$
and distribution function $\Psi({\bf v'})$, over a test star with
the same mass and distribution function $f({\bf w},t)$.

The assumption (\ref{fcampo}) is provided by the hydrodynamical
approximation, which establishes that the system is practically
space independent over the correlation range $\alpha$. Moreover, as
it is usual, we demand that the homogeneous distribution $\Psi$ must
satisfy the relation
\begin{equation}\label{normPhi}
    \int d^{3}{\bf v'}\Psi({\bf v'})=\frac{N}{V},
\end{equation}
where $V$ is the system's volume.

All the above considerations lead us to establish that, in order to
obtain the FPR equation starting from BBGKY hierarchy, we can choose
$g_{3}=g_{4}=\cdots=0$ and a two-point correlation function of the
form
\begin{equation}\label{g2wcg3}
    g_{2}({\bf w},{\bf w'},t)= \mathbf{\mathcal{G}}({\bf w},{\bf w'})\cdot\nabla_{vv'}
    [\Psi({\bf v'})f({\bf w},t)],
\end{equation}
where $\mathbf{\mathcal{G}}$ is given by the equation (\ref{Gcal}),
defining previously the constant $V_{o}$ through (\ref{beta}). We
have to point out that this model of correlation function, in the
context of a BBGKY scheme, is in close concordance with models based
on scattering-cross-section considerations as well as approaches
stated on a master equation (stochastic pictures).

\subsection{The Fokker-Planck-Rusenbluth Equation in 1PN
Approximation \label{1PNcollvlas}}

At this stage we can derive an alternative closed kinetic equation
for the self-gravitating system in 1PN approximation. Here, an
important fact, proceeding from our experience of sect.
\ref{covariant},  will play a crucial role: since in the 1PN
approach the momentum conservation for a system of point particles
is satisfied if and only if each particle obeys the Newtonian
equation of motion, the scattering process is characterized by a
differential cross section associated to the Newtonian law of force.
In other words: the statistical correlation between two
\emph{colliding} particles, at 1PN approximation level, must be
described by a two point correlation function corresponding to the
Newtonian scheme. This means that, as a consequence of the
statements established in the previous section, we can model
satisfactorily the relaxation process of the self-gravitating system
by introducing (\ref{g2wcg3}) in the collision term (\ref{F}), where
we also have to set $g_{3}=0$.

Then, by introducing (\ref{g2wcg3}) in (\ref{boltzmann}) and setting
$f'\approx\Psi({\bf v'})$, $f''\approx\Psi({\bf v''})$ (assumption
(\ref{fcampo})) in the collision term (\ref{F}), we obtain (see
\ref{ap3})
\begin{eqnarray}\label{vlasov2}
\frac{\partial f}{\partial t} + {\bf v} \cdot \frac{\partial
f}{\partial {\bf x}} + {\bf g} \cdot \frac{\partial f}{\partial {\bf
v}} + {\bf \Gamma}\cdot \frac{\partial f}{\partial {\bf v}} =
-\frac{\partial}{\partial
   v_{i}}(A^{i}f)+\frac{1}{2}\frac{\partial^{2}}{\partial v_{i}\partial
   v_{j}}(B^{ij}f),\nonumber\\
\end{eqnarray}
where $A^{i}$ and $B^{ij}$ are the 1PN-drift vector and the
1PN-diffusion tensor, respectively. They are given by the relations
\begin{eqnarray}\label{APN1}
   A^{i}&=&8\pi G^{2}m^{2}\ln \Lambda\frac{\partial}{\partial
   v_{i}}\int d^{3}{\bf v'}\frac{\Psi({\bf v'})}{|{\bf v}-{\bf
   v'}|}
   \nonumber\\&&\nonumber\\
   && +\int d^{3}{\bf v'}\Psi({\bf v'})\left[\frac{\partial}{\partial v'_{j}}-\frac{\partial}{\partial
    v_{j}}\right]\Omega^{ij}_{PN}({\bf w},{\bf v'}) \\
   &  & \nonumber\\
   &  & +\int d^{3}{\bf v'}d^{3}{\bf v''}\Psi({\bf v'})\Psi({\bf v''})\left[\frac{\partial}{\partial v''_{j}}-\frac{\partial}{\partial
    v'_{j}}\right]\Phi^{ij}_{PN}({\bf x},{\bf v'},{\bf v''}),\nonumber\\
    & &\nonumber\\
    B^{ij}&=&4\pi G^{2}m^{2}\ln \Lambda\frac{\partial^{2}}{\partial
   v_{i}\partial
   v_{j}}\int d^{3}{\bf v'}\Psi({\bf v'})|{\bf v}-{\bf v'}|\nonumber\\
    & &\nonumber\\
    &&-2\int d^{3}{\bf v'}\Psi({\bf v'})\Omega^{ij}_{PN}({\bf w},{\bf v'}),\label{BPN1}
\end{eqnarray}
where
\begin{eqnarray}\label{omegaPN}\label{PhiPN}
    \Omega^{ij}_{PN}({\bf w},{\bf v'})&=&\int d^{3}{\bf x'}\Lambda^{i}({\bf w},{\bf w'}){\cal G}^{j}({\bf w},{\bf w'})\\
    &&+2\frac{N}{V}\int d^{3}{\bf x'}d^{3}{\bf x''}\Upsilon^{i}_{S}({\bf x},{\bf x'},{\bf
    x''}){\cal G}^{j}({\bf w},{\bf w'}),\nonumber\\
    & &\nonumber\\
    \Phi^{ij}_{PN}({\bf x},{\bf v'},{\bf v''})&=&\int d^{3}{\bf x'}d^{3}{\bf x''}\Upsilon^{i}_{S}({\bf x},{\bf x'},{\bf
    x''}){\cal G}^{j}({\bf w'},{\bf w''}).
\end{eqnarray}
The first term on the r.h.s. of (\ref{APN1}) and (\ref{BPN1})
corresponds to the Newtonian contribution (classical Rosenbluth
coefficients) and the remainder is the post-Newtonian contribution
that can be calculated explicitly once we choose a particular
expression for the field star distribution $\Psi$. In this picture,
the post-Newtonian diffusion coefficients look like much more
involved than  in the formulation showed in sect. \ref{covariant}.
In this subject, we have to point out that they do not necessarily
coincide, even vanishing ternary contributions. The form in which
$A^{i}_{PN}$ and $B^{ij}_{PN}$ determines the collision term in the
first formulation, differs from the present one.

\section{Concluding Remarks \label{conclu}}

We have obtained two formulations of FPR-type equations that can be
used to model the evolution, in a diffusion approximation, of
stellar systems where the relativistic effects play a significant
role. In the first formulation the contribution of third order
correlation patterns to the diffusion coefficients is vanished,
while in the second one this fact is took into account. Since the
l.h.s of (\ref{fp1PN}) and  (\ref{vlasov2}) are equivalent, one can
expect they be consistent with the post-Newtonian equations of
hydrodynamics in general relativity \cite{rezania}. However the
generalization of the Eulerian equations of Newtonian hydrodynamics,
derived from each approach, will differ in the term associated to a
drift acceleration, since it will be determined by the diffusion
coefficients.

Since both (\ref{fp1PN}) and (\ref{vlasov2}) are expressed in a
non-manifestly covariant fashion (as a consequence of the 1PN
scheme) its physical interpretation, as well as their differences
with the usual FPR equation, can be elucidated from a classical
Newtonian point of view. We have to point out that they have a
common point: Once we choose an adequate field star distribution
function, the diffusion coefficients (\ref{Ai2PN})-(\ref{Bij2PN}) or
(\ref{APN1})-(\ref{BPN1}) can be calculated explicitly, making
possible to find numerical solutions of (\ref{fp1PN}) and
(\ref{vlasov2}) (for example, using the formalism showed in
\cite{letelier}).

 The KD formalism allows us to derive, in a relatively simple way,
the first 1PN-BBGKY relation, eq. (\ref{boltzmann}). It constitutes
an interesting example about the importance of this method, when we
deal with the statistical description of dynamical systems described
by velocity-dependent interactions (see \cite{golds}). Perhaps the
most relevant fact, that can be noticed through (\ref{boltzmann}),
is the apparition of three order correlation patterns in the
collision term. This feature, absent in the purely newtonian case,
is a consequence of the ternary interaction ${\bf \Upsilon}$ (see
eq. (\ref{Upsilon})). One could go far away and think that more
accurate post Newtonian approximations (2PN, 3PN and so on) would
lead to descriptions based on correlations of greater order. It
would be not a surprising fact that the Einstein's gravity theory
\emph{increases the correlation level}, in a non-equilibrium
statistical mechanics description.

Assumptions (\ref{g2wcg3}) and (\ref{fcampo}), reduce
(\ref{boltzmann}) (in the purely Newtonian case) to the usual FPR
equation, (\ref{fokk-planck}). This fact suggests that the local
approximation, employed in the stochastic approaches of kinetic
theory (see for example \S 8.3 of \cite{BinTre}), is equivalent to
the hydrodynamic-WCG approximation, used here. Through such
approximation we model the star cluster relaxation process by the
correlation dynamics corresponding to a gas dominated by attractive
short range interactions, $V(r)=-(2Gm\ln
\Lambda/\alpha\sqrt{\pi})\exp(-r^{2}/\alpha^{2})$. The interaction
range $\alpha$, introduced here, defines a characteristic scale of
distances over which (i) encounters play a dominant role, and (ii)
the distribution function can be considered homogeneous. One can
estimate the magnitude of such length, taking into account the
considerations used in the local approximation, and say that
$\alpha$ must be of the order of the minimum impact parameter for
which such approximation holds. That is, $\alpha\sim R/N$, where $R$
is the system's characteristic radius \cite{BinTre}. According to
these statements, the equation (\ref{vlasov2}), showed in sect.
\ref{1PNcollvlas}, describes the evolution of the distribution
function corresponding to a typical test particle interacting with
an homogeneous sea of test stars in thermodynamic equilibrium,
taking into account 1PN corrections.

\begin{appendix}

\section{Post Newtonian potentials for $N$ point-like particles \label{ap1}}

In the 1PN approximation, $\Phi$, ${\cal\xi}$ and $\psi$ are given
by \cite{Wein}
\begin{eqnarray} \Phi ({\bf x},t)&=& -G \int d^{3}{\bf x}' \frac{\stackrel{0 \: \:
\: \: }{T^{00}}({\bf x}',t)}{|{\bf x}-{\bf x}'|},\label{apafi}
\\ && \nonumber\\
 \psi ({\bf x},t) &=& - \int \frac{d^{3}{\bf x}'}{ |{\bf x}-{\bf
x}'|} \left [ \frac{1}{4\pi}\frac{\partial^{2}\Phi ({\bf
x}',t)}{\partial t^{2}} \right. \left. + G\stackrel{2 \: \: \: \:
}{T^{00}}({\bf x}',t) + G\stackrel{2 \: \: \: \: }{T^{aa}}({\bf
x}',t) \right ], \label{apasi}\nonumber \\ & &
\\
{\cal\xi} _{a} ({\bf x},t)&=& -4G \int d^{3}{\bf x}'
\frac{\stackrel{1 \: \: \: \: }{T^{a0}}({\bf x}',t)}{|{\bf x}-{\bf
x}'|}.\label{apaxi}
\end{eqnarray}
(We denote $a,b=1,2,3$). For a system composed by $N$ identical
point-like particles, the energy-momentum tensor components, at this
order, are
\begin{eqnarray}
  \stackrel{0 \: \:
\: \: }{T^{00}}({\bf x}',t)&=& m\sum_{i=1}^{N}\delta({\bf x}'-{\bf
x}_{i}(t)),\label{apt000}
\\ && \nonumber\\
  \stackrel{2 \: \:
\: \: }{T^{00}}({\bf x}',t)&=& m\sum_{i=1}^{N}\left[\Phi({\bf
x}',t)+\frac{{\bf v}_{i}^{2}}{2}\right]\delta({\bf x}'-{\bf
x}_{i}(t)),\label{apt002}
\\ && \nonumber\\
  \stackrel{1 \: \:
\: \: }{T^{0a}}({\bf x}',t)&=&m\sum_{i=1}^{N}v^{a}_{i}\delta({\bf
x}'-{\bf x}_{i}(t)),\label{apt0j1}
\\ && \nonumber\\
  \stackrel{2 \: \:
\: \: }{T^{ab}}({\bf
x}',t)&=&m\sum_{i=1}^{N}v_{i}^{a}v_{i}^{b}\delta({\bf x}'-{\bf
x}_{i}(t)).\label{aptjk2}
\end{eqnarray}
Introducing (\ref{apt000}) in (\ref{apafi}), and (\ref{apt0j1}) in
(\ref{apaxi}), we obtain
\begin{equation} \Phi({\bf x},t) = -\sum_{i=1}^{N}
\frac{Gm}{|{\bf x}-{\bf x}_{i}(t)|},\label{apfi}
\end{equation}
\begin{equation}
{\cal\xi}({\bf x},t)=-\sum_{i=1}^{N}\frac{4Gm{\bf v}_{i}}{|{\bf
x}-{\bf x}_{i}(t)|},\label{apxi}
\end{equation}
while, introducing (\ref{apt002}) and (\ref{aptjk2}) in
(\ref{apasi}) (expression $T^{aa}$ indicates sum over $a$), $\psi$
takes the form
\begin{eqnarray}\label{apasi1}
 \psi({\bf x},t)& = &
-\sum_{i=1}^{N}\frac{(3/2)Gm{\bf v}_{i}^{2}}{|{\bf x}-{\bf x}_{i}|}
+\sum_{i=1}^{N}\sum_{j\neq i} \frac{G^{2}m^{2}}{|{\bf x}-{\bf
x}_{i}||{\bf x}_{i}-{\bf
x}_{j}|}\nonumber\\
   &  & +
\frac{Gm}{4\pi}\frac{\partial^{2}}{\partial t^{2}}\sum_{i=1}^{N}\int
\frac{d^{3}{\bf x}'}{|{\bf x}-{\bf x}'||{\bf x}'-{\bf x}_{i}(t)|}.
\end{eqnarray}
In order to solve the integral on the right hand we introduce the
identity \cite{kn:1}
$$
\frac{1}{|{\bf x}-{\bf x}'|}=\frac{1}{2\pi^{2}}\int \frac{d^{3}{\bf
k}}{k^{2}}e^{i{\bf k}\cdot({\bf x}-{\bf x}')},
$$
that permits to write
$$
\frac{1}{|{\bf x}-{\bf x}'||{\bf x}'-{\bf
x}_{i}|}=\frac{1}{4\pi^{4}}\int \frac{d^{3}{\bf k}d^{3}{\bf
k}'}{k^{2}k'^{2}}e^{i({\bf k}\cdot{\bf x}-{\bf k}'\cdot{\bf
x}_{i})}e^{i({\bf k}'-{\bf k})\cdot{\bf x}'}.
$$
The integral over ${\bf x}'$ of this last expression, in virtue of
the identity
$$
\delta({\bf k}'-{\bf k})=\frac{1}{(2\pi)^{3}}\int d^{3}{\bf
x}'e^{i({\bf k}'-{\bf k})\cdot{\bf x}'},
$$
equals to
\begin{equation}\label{apint1}
    \int\frac{d^{3}{\bf
x}'}{|{\bf x}-{\bf x}'||{\bf x}'-{\bf x}_{i}|}=\frac{2}{\pi}\int
\frac{d^{3}{\bf k}}{k^{4}}e^{i{\bf k}\cdot({\bf x}-{\bf x}_{i})}.
\end{equation}
Defining cartesian coordinates, such that $k_{z}$ is in the
direction of
\begin{equation}\label{apar}
    {\bf r}={\bf x}-{\bf x}_{i},
\end{equation}
and then, changing to spherical polar coordinates and introducing
the transformation $u=kr=k|{\bf x}-{\bf x}_{i}|$, the right hand
side integral of (\ref{apint1}) takes the form
\begin{equation}\label{apint2}
    \int
\frac{d^{3}{\bf k}}{k^{4}}e^{i{\bf k}\cdot({\bf x}-{\bf
x}_{i})}=4\pi |{\bf x}-{\bf x}_{i}|\int_{0}^{\infty}du\frac{\sin
u}{u^{3}}.
\end{equation}
Since
$$
\int_{0}^{\infty}du\frac{\sin u}{u^{3}}=-\frac{\pi}{4},
$$
we find that (\ref{apint1}) reduces to
\begin{equation}\label{apint3}
     \int\frac{d^{3}{\bf
x}'}{|{\bf x}-{\bf x}'||{\bf x}'-{\bf x}_{i}|}=-2\pi|{\bf x}-{\bf
x}_{i}|.
\end{equation}
Then, we have to take into account the term
$$
\frac{\partial^{2}}{\partial t^{2}}|{\bf x}-{\bf
x}_{i}(t)|=\frac{{\bf v}_{i}^{2}}{|{\bf x}-{\bf x}_{i}|}-\frac{[{\bf
v}_{i}\cdot({\bf x}-{\bf x}_{i})]^{2}}{|{\bf x}-{\bf
x}_{i}|^{3}}-\frac{{\bf x}-{\bf x}_{i}}{|{\bf x}-{\bf
x}_{i}|}\cdot\frac{d{\bf v}_{i}}{dt}.
$$
in (\ref{apasi1}). Here, in agreement with the order of
approximation, we must to take
$$
\frac{d{\bf v}_{i}}{dt}=-Gm\sum_{j\neq i}\frac{{\bf x}_{i}-{\bf
x}_{j}}{|{\bf x}_{i}-{\bf x}_{j}|^{3}}.
$$
Finally, we can write (\ref{apasi1}) as
\begin{eqnarray}
\psi({\bf x},t) &=& - G^{2} m^{2} \sum_{i=1}^{N} \sum_{j\neq i}
\left\{ \frac{({\bf x} - 3 {\bf x}_{i} + 2 {\bf x}_{j}) \cdot ({\bf
x}_{i} - {\bf x}_{j})}{2 |{\bf x} - {\bf x}_{i}| |{\bf x}_{i} - {\bf
x}_{j}|^{3}} \right\}
\nonumber \\
&& \nonumber \\
&&- G m \sum_{i=1}^{N} \left\{\frac{2 {\bf v}_{i}^{2}}{|{\bf x} -
{\bf x}_{i}|} - \frac{[{\bf v}_{i} \cdot ({\bf x} - {\bf
x}_{i})]^{2}}{2 |{\bf x} - {\bf x}_{i}|^{3}} \right\} .\label{apsi}
\end{eqnarray}
By introducing relations (\ref{apfi}), (\ref{apxi}) and (\ref{apsi})
in (\ref{gama}), we can write it as (\ref{gama2}).

\section{Derivation of relations (\ref{FN2})-(\ref{B2})\label{ap2}}

Introducing (\ref{g2wcg2}) in (\ref{FN1}), it can be cast as
\begin{equation}\label{apFN1}
   -\frac{\partial}{\partial{\bf v}}\cdot{\bf F}_{N}=\int d^{3}{\bf v'}\nabla_{vv'}\cdot{\bf
   \Omega}_{N}\cdot\nabla_{vv'}[f({\bf x},{\bf v'},t)f({\bf x},{\bf
   v},t)],
\end{equation}
where the subscript $N$ indicates the newtonian case and ${\bf
\Omega}_{N}$ is a second rank tensor, defined as
\begin{equation}\label{apomega1}
    {\bf \Omega}_{N}=\int_{0}^{\infty}d\tau\int d^{3}{\bf r}
   \left[\frac{\partial \Phi({\bf r})}{\partial{\bf x}}\right]\left[\frac{\partial V({\bf r}-\tau{\bf u}_{vv'})}{\partial{\bf
   x}}\right],
\end{equation}
where we have changed the integration domain ${\bf x'}$ by ${\bf
r}={\bf x}-{\bf x'}$, and called $\Phi({\bf r})=-Gm/r$. It is
convenient to express $\Phi({\bf r})$ and $V({\bf r})$ in the
Fourier expansion:
\begin{equation}\label{apfourier}
   \Phi({\bf r})=\int d^{3}{\bf k}\Phi_{k}e^{i{\bf k}\cdot{\bf
   r}},\:\:\:\:\:\:\:\:V({\bf r})=\int d^{3}{\bf k}V_{k}e^{i{\bf k}\cdot{\bf
   r}},
\end{equation}
with
\begin{equation}\label{aptranfourier}
    \Phi_{k}=-\frac{Gm}{2\pi^{2}k^{2}},\:\:\:\:\:\:\:\:V_{k}=\frac{\alpha^{3}V_{o}}{8\pi^{3/2}}e^{-(k\alpha
    /2)^{2}}.
\end{equation}
Introducing (\ref{apfourier}) in (\ref{apomega1}), we find
\begin{equation}\label{apomega2}
    {\bf \Omega}_{N}=8\pi^{3}\int_{0}^{\infty}d\tau\int d^{3}{\bf k}e^{i\tau{\bf k}\cdot{\bf u}_{vv'}}\Phi_{k}V_{k}{\bf k}{\bf
    k}.
\end{equation}
The integral with respect to $\tau$ can be evaluated using the
representation (see appendix 2 of \cite{bal2})
\begin{equation}\label{apsingfunc}
    \int_{0}^{\infty}d\tau e^{\pm ix\tau}=\pi\delta(x)\pm i{\cal
    P}\left(\frac{1}{x}\right),
\end{equation}
where ${\cal P}(1/x)$ denotes the principal part. Taking into
account that $\delta(x)$ is an even function, ${\cal P}(1/x)$ is
odd, and $\Phi_{k}V_{k}{\bf k}{\bf k}$ is even in the vector ${\bf
k}$, we obtain
\begin{equation}\label{apomega2}
    {\bf \Omega}_{N}=8\pi^{4}\int d^{3}{\bf k}\delta({\bf k}\cdot{\bf u}_{vv'})\Phi_{k}V_{k}{\bf k}{\bf
    k}.
\end{equation}
The calculation of the above integral is simplified using spherical
coordinates and choosing the $Z$-axis in the ${\bf u}_{vv'}$
direction (details of this transformation are shown in \cite{bal},
sect 11.6). We find that
\begin{equation}\label{apomega3}
    {\bf \Omega}_{N}=C\frac{u_{vv'}^{2}{\bf I}-{\bf u}_{vv'}{\bf u}_{vv'}}{u_{vv'}^{3}}
\end{equation}
where ${\bf I}$ is the identity second rank tensor and
\begin{equation}\label{apC}
    C=8\pi^{5}\int_{0}^{\infty}dk k^{3}\Phi_{k}V_{k}=-\pi^{3/2}
    Gm\alpha V_{o}.
\end{equation}
The relation (\ref{apomega3}) allows us to write the velocity
divergence of ${\bf F}_{N}$ in the Fokker-Planck form:
\begin{equation}\label{apapFN2}
   -\frac{\partial}{\partial{\bf v}}\cdot{\bf F}_{N}=-\frac{\partial}{\partial
   v_{i}}(\tilde{A}^{i}_{N}f)+\frac{1}{2}\frac{\partial^{2}}{\partial v_{i}\partial
   v_{j}}(\tilde{B}^{ij}_{N}f)
\end{equation}
where we have used the sum convention ($i,j=1,2,3$) and
\begin{eqnarray}
    \tilde{A}^{i}_{N}&=&\int d^{3}{\bf v'}f({\bf x},{\bf v'})\left[\frac{\partial}{\partial v_{j}}-\frac{\partial}{\partial
    v'_{j}}\right]\Omega^{ij}_{N},\label{apA1}\\
    &&\nonumber\\
    \tilde{B}^{ij}_{N}&=&2\int d^{3}{\bf v'}f({\bf x},{\bf v'})\Omega^{ij}_{N}.\label{apB1}
\end{eqnarray}
Finally, introducing (\ref{apomega3}) and (\ref{apC}) in the last
relations, we obtain (\ref{A2})-(\ref{B2}).

\section{Derivation of the post Newtonian diffusion coefficients\label{ap3}}

It is possible to write the remaining part of the collision term in
a Fokker-Planck form, similar to relation (\ref{FN2}). We first
write ${\bf F}={\bf F}_{N}+{\bf F}_{PN}$, considering only ${\bf
F}_{PN}$ (the \emph{post Newtonian part} of ${\bf F}$). From
(\ref{F}) we note that a simplification is introduced if we define
symmetric and antisymmetric (with respect to ${\bf x'},{\bf x''}$)
fields
\begin{equation}\label{apsimUps}
    {\bf \Upsilon}_{S}({\bf x},{\bf x'},{\bf x''})\equiv
    \frac{1}{2}[{\bf \Upsilon}({\bf x},{\bf x'},{\bf x''})+{\bf \Upsilon}({\bf x},{\bf x''},{\bf
    x'})],
\end{equation}
\begin{equation}\label{apsimUps}
    {\bf \Upsilon}_{A}({\bf x},{\bf x'},{\bf x''})\equiv
    \frac{1}{2}[{\bf \Upsilon}({\bf x},{\bf x'},{\bf x''})-{\bf \Upsilon}({\bf x},{\bf x''},{\bf
    x'})].
\end{equation}
With the help of these definitions, remembering that $g_{2}$ is
symmetric and neglecting $g_{3}$, ${\bf F}_{PN}$ can be cast as
\begin{eqnarray}\label{apFPN}
{\bf F}_{PN} & = & \int d^{6}{\bf w'}{\bf
\Lambda}({\bf w},{\bf w'}) g_{2}({\bf w},{\bf w'})  \nonumber\\
&&  \nonumber   \\
& +& \int d^{6}{\bf w'} d^{6}{\bf w''}{\bf \Upsilon}_{S}({\bf
x},{\bf x'},{\bf x''})\left[ g_{2}({\bf w'},{\bf w''}) f +
2g_{2}({\bf w},{\bf w'}) f'' \right].
\end{eqnarray}
Introducing (\ref{g2wcg3}) in this relation, taking into account the
assumption (\ref{fcampo}), by which one can set $f'\approx\Psi({\bf
v'})$, $f''\approx\Psi({\bf v''})$  in the collision term, we find
that the velocity divergence of the resulting expression can be
written as
\begin{equation}\label{apFPN2}
   -\frac{\partial}{\partial{\bf v}}\cdot{\bf F}_{PN}=-\frac{\partial}{\partial
   v_{i}}(A^{i}_{PN}f)+\frac{1}{2}\frac{\partial^{2}}{\partial v_{i}\partial
   v_{j}}(B^{ij}_{PN}f).
\end{equation}
Here we have introduced the \emph{post Newtonian diffusion
coefficients}:
\begin{eqnarray}\label{apAPN1}
   A^{i}_{PN} & = & \int d^{3}{\bf v'}\Psi({\bf v'})\left[\frac{\partial}{\partial v'_{j}}-\frac{\partial}{\partial
    v_{j}}\right]\Omega^{ij}_{PN} \\
   &  & \nonumber\\
   &  & +\int d^{3}{\bf v'}d^{3}{\bf v''}\Psi({\bf v'})\Psi({\bf v''})\left[\frac{\partial}{\partial v''_{j}}-\frac{\partial}{\partial
    v'_{j}}\right]\Phi^{ij}_{PN},\nonumber\\
    & &\nonumber\\
    B^{ij}_{PN}&=&-2\int d^{3}{\bf v'}\Psi({\bf v'})\Omega^{ij}_{PN},\label{apBPN1}
\end{eqnarray}
where
\begin{eqnarray}\label{apomegaPN}\label{apPhiPN}
    \Omega^{ij}_{PN}&=&\int d^{3}{\bf x'}\Lambda^{i}({\bf w},{\bf w'}){\cal G}^{j}({\bf w},{\bf w'})\\
    &&+2\frac{N}{V}\int d^{3}{\bf x'}d^{3}{\bf x''}\Upsilon^{i}_{S}({\bf x},{\bf x'},{\bf
    x''}){\cal G}^{j}({\bf w},{\bf w'}),\nonumber\\
    & &\nonumber\\
    \Phi^{ij}_{PN}&=&\int d^{3}{\bf x'}d^{3}{\bf x''}\Upsilon^{i}_{S}({\bf x},{\bf x'},{\bf
    x''}){\cal G}^{j}({\bf w'},{\bf w''}).
\end{eqnarray}
We obtain (\ref{vlasov2}) adding (\ref{apFPN2}) and (\ref{apapFN2})
(choosing $V_{o}$ according to (\ref{beta})).

\end{appendix}


\section*{References}

\begin{thebibliography}{20}
%%%%%%%%%%%%%%%%%%%%%%

\bibitem{rezania} V. Rezania1 and Y. Sobouti. Astron.Astrophys. 354, 1110–1114 (2000)

\bibitem{rees} M. J. Rees. Annual Reviews of Astronomy and Astrophysycs. 22, 471 (1984)

\bibitem{shapiro} S. L. Shapiro and S. A. Teukolsky. Astroph. J. Lett. 292, 41 (1985)

\bibitem{BinTre} J. Binney and S. Tremaine. {\em Galactic Dynamics}. Princeton
University Press (1987).

\bibitem{rosenbluth} M. N. Rosenbluth, W. M. MacDonald and D.L. Judd. Phys. Rev. 107 (1), 1-6 (1957).

\bibitem{chan1} S. Chandrasekar. Reviews of Modern Physics \textbf{21}(3),
383-388 (1949).

\bibitem{chan2} S. Chandrasekar. Reviews of Modern Physics \textbf{15}(1),
1-89 (1943).

\bibitem{chacon} G. Chac\'on-Acosta and G. M. Kremer. Phys. Rew. E \textbf{76},
021201 (2007).

\bibitem{kand1} W. Israel and H. Kandrup. Annals of Physics \textbf{152}(1), 30-84
(1984)

\bibitem{kand2} H. Kandrup.  Annals of Physics \textbf{153}(1), 44-102 (1984)

\bibitem{kand3} H. Kandrup. Annals of Physics \textbf{169}(2), 352-413 (1986).

\bibitem{kremer} C. Cercignani and G. M. Kremer.
\emph{The relativistic Boltzmann Equation: Theory and Applications}
(Birkh\"{a}user Verlag, Basel, 2002).

\bibitem{landau} L. D. Landau. Phys. Z. Sowj. Union. \textbf{10}, 154 (1936).

\bibitem{bal} R. Balescu. {\em Equilibrium and Non-Equilibrium Statistical
Mechanics}. Wiley Interscience (1975).

\bibitem{Kunk} W Kunkel. {\em Plasma Physics in Theory and
Application}. Mc Graw-Hill (1966).

\bibitem{Bogoliubov} N. N. Bogoliubov . J. Phys. USSR.  \textbf{10}, 257-265 (1938).

\bibitem{born-green} M. Born. and H.S. Green. Proc. Roy. Soc. Lond. A188, \textbf{10} (1946).

\bibitem{kirkwood} J. G. Kirkwood. J. Chem. Phys.  \textbf{14}, 180(1946).

\bibitem{yvon} J. Yvon. {\em Les Corr\'elations et l'Entropie en M\'ecanique Statistique Classique}.  Dunod, Par\'\i s (1935)

\bibitem{Klim} Y. L. Klimontovich. {\em The Statistical Theory of Non-Equilibrium Processes in a Plasma}. Cambridge: MIT Press (1967).

\bibitem{Dup} T. H. Dupree. Phys. Fluids, \textbf{10}, 1049 (1967).

\bibitem{Wein} S. Weinberg. {\em Gravitation and Cosmology}. John Wiley (1972).

\bibitem{zim} W. Zimdahl. Class. Quantum Grav. \textbf{6}, 1879-1892 (1989).

\bibitem{golds} P. Goldstein and L.A. Turski. Physica \textbf{89A}, 481-500(1977).

\bibitem{MaBer} C. P. Ma  and E. Bertschinger. Ap. J. \textbf{612}, 28-49 (2004).

\bibitem{vlasov} A. Vlassov. Zh. Eksp. Teor. Fiz.  \textbf{8}, 481-500(1938).

\bibitem{golse} F. Golse. Journ´ees ´ Equations aux d´eriv´ees
partielles, Forges-les-Eaux, 2–6 Juin (2003).

\bibitem{letelier} M. Ujevic and P. Letelier. J. Comp. Phys. 215, 485–505 (2006)

\bibitem{kn:1} G. Arfkeen. {\em Mathematical Methods for Physicists}. Academic
Press. Third edition (1985).

\bibitem{bal2} R. Balescu. {\em Statistical Mechanics of Charged
Particles}. Wiley Interscience (1963).







%\bibitem{fokker} L. Onsager and S. Machlup. Phys. Rev.  \textbf{91}, 1505, 1512 (1953)

%\bibitem{Prig} I. Prigogine. {\em Non-Equilibrium Statistical
%Mechanics}. Interscience Publishers (1962).

%\bibitem{kand5} H. Kandrup. Physica A \textbf{126}(3), 461-473 (1984).

%\bibitem{andreason}H. Andr\'easson.
%Living Rev. Relativity \textbf{8} (2), 1-33  [Online Article:
%http://www.livingreviews.org/lrr-2005-2] (2005).

%\bibitem{cluster}M. Bottaccio, L. Pietroneroa, A. Amicia, P. Miocchia, R. C.
%Dolcettaa and M. Montuorib. Physica A, \textbf{305} (1-2), 247-252
%(2002).

%\bibitem{kand4} H. Kandrup. Physics Reports, Phys Lett. \textbf{63} (1), 1-59 (1980).


%\bibitem{Kukharenko} Y. Kukharenko, A. Vityazev and A. Bashkirov. Phys Let
%A \textbf{195} (1), 27-30 (1994)

%\bibitem{Haggerty} M. J. Haggerty. Physica \textbf{50} (3), 391-396 (1970)

%\bibitem{blanchet} L. Blanchet. Living Rev. Relativity
%5 (3), 1-81 [Online Article:
%http://www.livingreviews.org/lrr-2002-3] (2002).







%%%%%%%%%%%%%%%%%%%%%%

\end{thebibliography}
\end{document}